\documentclass[conference]{IEEEtran}
\IEEEoverridecommandlockouts

\usepackage{cite}
\usepackage{amsmath,amssymb,amsfonts,amsthm}
\usepackage{algorithmic}
\usepackage{textcomp}
\usepackage{multirow}
\usepackage{graphicx}
\usepackage{url}
\usepackage{subcaption}
\usepackage{xspace}

\usepackage{enumitem}
\usepackage{hyperref}
\usepackage{cleveref}
\usepackage{fancyvrb}
\usepackage{framed}

\usepackage[framemethod=TikZ]{mdframed}

\usepackage{pgfplots}
\usepackage{pgfplotstable}
\pgfplotsset{compat=1.18}

\usepackage{booktabs}
\usepackage{caption}
\usepackage{adjustbox}
\usepackage{listings}[language=Java]

\def\BibTeX{{\rm B\kern-.05em{\sc i\kern-.025em b}\kern-.08em
    T\kern-.1667em\lower.7ex\hbox{E}\kern-.125emX}}

\definecolor{bg}{rgb}{0.98, 0.98, 0.98}

\usepackage{tcolorbox}
\definecolor{bg}{gray}{0.95}
\definecolor{darkred}{RGB}{139, 0, 0} %

\colorlet{shadecolor}{gray!30}
\definecolor{Gray}{gray}{0.8}

\newenvironment{packed_itemize}{
\begin{list}{\labelitemi}{\leftmargin=1.0em}
 \setlength{\itemsep}{2.5pt}
 \setlength{\parskip}{0pt}
 \setlength{\parsep}{0pt}
 \setlength{\headsep}{0pt}
 \setlength{\topskip}{0pt}
 \setlength{\topmargin}{0pt}
 \setlength{\topsep}{0pt}
 \setlength{\partopsep}{0pt}
}{\end{list}}

\begin{document}

\title{An Empirical Study of Production Incidents in Generative AI Cloud Services}

\author{
    \IEEEauthorblockN{Haoran Yan$^{*1}$, Yinfang Chen$^{* 2}$, Minghua Ma$^{\dag3}$, Ming Wen$^{1}$, Shan Lu$^{3}$, Shenglin Zhang$^{4}$, Tianyin Xu$^{2}$}
    \IEEEauthorblockN{Rujia Wang$^{3}$, Chetan Bansal$^{3}$, Saravan Rajmohan$^{3}$, Qingwei Lin$^{5}$, Chaoyun Zhang$^{5}$, Dongmei Zhang$^{5}$}
    \IEEEauthorblockN{}
    \IEEEauthorblockA{$^1$Huazhong University of Science and Technology, China, \{haoran\_yan, mwenaa\}@hust.edu.cn}
    \IEEEauthorblockA{$^2$University of Illinois Urbana-Champaign, USA, \{yinfang3, tyxu\}@illinois.edu}
    \IEEEauthorblockA{$^3$Microsoft, USA, \{minghuama, shanlu, rujiawang, chetanb, saravan.rajmohan\}@microsoft.com}
    \IEEEauthorblockA{$^4$Nankai University, China, \{zhangsl\}@nankai.edu.cn}
    \IEEEauthorblockA{$^5$Microsoft, China, \{chaoyun.zhang, qlin, dongmeiz\}@microsoft.com}
    \thanks{*~Haoran Yan and Yinfang Chen contributed equally. Work was performed during their internship at Microsoft.}
    \thanks{\dag~Corresponding author.}
}

\newcounter{insightC}
\newcounter{caseC}

\def\caseCautorefname{Case}
\def\insightCautorefname{Finding}

\newcommand{\block}[1]{\textcolor{blue}{$\qquad$#1}}
\newcommand{\new}[1]{\textcolor{black}{#1}}

\setlength{\FrameSep}{3pt}

\newtheoremstyle{compactdef}
  {0pt}     
  {0pt}     
  {}        
  {0pt}     
  {}
  {:}       
  { }       
  {}        
\theoremstyle{compactdef}
\newtheorem{findinner}{\textit{\textbf{Finding}}}

\newenvironment{find}
  {\begin{shaded}\begin{findinner}}
  {\end{findinner}\end{shaded}}

\newcommand{\insight}[1]{

  \begin{find}
    #1
  \end{find}
}

\newcommand{\case}[4]{
    \refstepcounter{caseC} %
    \begin{center}
    \begin{tcolorbox}[
                    colback=bg, 
                    colframe=black, 
                    width=8.5cm, 
                    arc=2mm, auto outer arc, 
                    boxrule=0.5pt,
                    left=1mm, 
                    right=1mm, 
                    top=1mm,
                    bottom=1mm
                ]
    \textbf{Case \arabic{caseC}:~}#1\\ %
    \textbf{Symptom:~}\new{#2}\\ %
    \textbf{Root Cause:~}\new{#3}\\ %
    \textbf{Mitigation:~}\new{#4} %
    \end{tcolorbox}
    \end{center}
}

\newcommand{\yfcase}[4]{
    \refstepcounter{caseC} %
    \begin{center}
    \begin{tcolorbox}[
                    width=\linewidth, 
                    title={\textbf{Case \arabic{caseC}:~}#1}, 
                    arc=2mm, 
                    auto outer arc, 
                    boxrule=0.5pt,
                    left=1mm, 
                    right=1mm, 
                    top=1mm,
                    bottom=1mm
                ]
    \textbf{Symptom:~}\new{\color{blue}#2}\\ %
    \textbf{Root Cause:~}\new{\color{blue}#3}\\ %
    \textbf{Mitigation:~}\new{\color{blue}#4} %
    \end{tcolorbox}
    \end{center}
}

\newcommand{\cyfcase}[3]{
    \refstepcounter{caseC} %
    \begin{center}
    \begin{tcolorbox}[
                    width=\linewidth, 
                    title={\small\textbf{Case \arabic{caseC}:~}#1}, 
                    arc=2mm, 
                    auto outer arc, 
                    boxrule=0.5pt,
                    left=1mm, 
                    right=1mm, 
                    top=1mm,
                    bottom=0.01mm
                ]
    \small{\textbf{Symptom:~}}\small\new{#2}\\ %
    \small{\textbf{Root Cause:~}}\small\new{#3} %
    \end{tcolorbox}
    \end{center}
}

\newcommand{\point}[1]{\vspace{0.45mm}\noindent\textbf{#1}.}
\newcommand{\mh}[1]{{\color{purple}[MH: #1]}}
\newcommand{\civi}[1]{{\color{purple}[Ming: #1]}}
\newcommand{\xiaohu}[1]{{\color{violet}[xiaohu: #1]}}
\newcommand{\haoran}[1]{{\color{blue}[haoran: #1]}}
\newcommand{\hr}[1]{{{\color{red} #1}}}
\newcommand{\yinfang}[1]{{{\color{red!85!black} [yinfang: #1]}}}
\newcommand{\yf}[1]{{{\color{blue} #1}}}
\newcommand{\gai}[0]{GenAI}
\newcommand{\cloud}[0]{\textit{Microsoft}} %
\newcommand{\aoai}[0]{\textit{Azure OpenAI}} %

\def\finding{\noindent\textit{Finding}~}
\def\ie{\textit{i.e.},~}
\def\etal{\textit{et al.}~}
\def\etc{\textit{etc.}~}
\def\eg{\textit{e.g.},~}
\def\Eg{\textit{E.g.},~}
\def\vs{\textit{vs.}~}
\def\aka{\textit{a.k.a.}~}
\def\Snospace~{\S{}}

\pagestyle{plain}

\maketitle

\begin{abstract}

The ever-increasing demand for generative artificial intelligence (GenAI) has motivated cloud-based GenAI services such as Azure OpenAI Service. Like any large-scale cloud service, failures are inevitable in cloud-based GenAI services, resulting in user dissatisfaction and significant monetary losses. However, GenAI cloud services, featured by their massive parameter scales, hardware demands, and usage patterns, present unique challenges, including generated content quality issues and privacy concerns, compared to traditional cloud services. To understand the production reliability of GenAI cloud services, we analyzed production incidents from Microsoft spanning in the past four years. Our study (1) presents the general characteristics of GenAI cloud service incidents at different stages of the incident life cycle; (2) identifies the symptoms and impacts of these incidents on GenAI cloud service quality and availability; (3) uncovers why these incidents occurred and how they were resolved; (4) discusses open research challenges in terms of incident detection, triage, and mitigation, and sheds light on potential solutions. 

\end{abstract}


\section{Introduction}

In recent years, there have been significant advancements in generative artificial intelligence (\gai{}), particularly in Large Language Models (LLMs) and their applications across various fields. Beyond natural language processing, these models have also shown new capabilities in image recognition~\cite{DBLP:journals/corr/abs-2403-01209, DBLP:journals/corr/abs-2309-03905}, data analysis\cite{zhang2024allhands, qiao2023taskweaver}, software engineering~\cite{jiang2023xpert, chen2024automatic, jin2023assess, zhang2024ufo}, and more~\cite{DBLP:journals/corr/abs-2309-11325, DBLP:journals/npjdm/RazaVK24, DBLP:journals/corr/abs-2311-07816}. The emergence of models like the GPT-4 family marks a new era, with capabilities extending to complex reasoning~\cite{DBLP:journals/corr/abs-2310-03965, DBLP:journals/corr/abs-2401-00125}, creative thinking~\cite{DBLP:journals/corr/abs-2401-12491, DBLP:journals/corr/abs-2310-06155, ruomeng2024thought}, and even surpassing human expertise in certain tasks~\cite{bubeck2023sparks}. This innovation has resulted in impactful research findings and practical applications with substantial implications for scientific research and socio-economic development.

The demands of \gai{} come with the requirements of unprecedented computational resources, including the hardware for operating the models as well as infrastructure systems for efficiently allocating and utilizing such resources \cite{miao2023towards}.
However, both the acquisition of the required resources and their efficient management pose significant challenges to individuals and even enterprises.
Therefore, it motivates the development of \textit{\gai{} cloud services}, which offer a platform where developers and users can create, deploy, and utilize large models without substantial hardware and software investments, e.g., Cloud for AI (Cloud4AI), and also incorporate model APIs within cloud systems, e.g., AI for Cloud (AI4Cloud) \cite{jiang2023xpert, chen2024automatic, jin2023assess, chen2024aiopslab, shetty2024building, yu2024monitorassistant}. 
Popular \gai{} cloud services include Azure OpenAI, Amazon Bedrock, IBM Watson, and Anthropic Claude.
\gai{} cloud services afford enterprises the infrastructure necessary for the deployment and maintenance of \gai{} models, based on which users can further interact, analyze, and fine-tune such models.
Moreover, they are crucial in promoting collaboration among researchers by providing shared access to advanced models and computational resources.

As with any large-scale cloud services, GenAI cloud services are not immune to occasional incidents. These events, while often unavoidable due to the complexity and scale of the systems involved, have the potential to impact user experience and, in some cases, result in challenges such as user dissatisfaction or economic implications.
For example, OpenAI recently experienced an incident where request failures and high latency severely impacted ChatGPT's API and functionalities~\cite{elevated_gpt}. 
However, despite the critical importance of reliability in \gai{} cloud services, there is a notable lack of research focusing on their reliability and incident management.
Therefore, understanding the characteristics of these incidents—including detection, triage, diagnosis, and mitigation—is crucial for enhancing the quality of \gai{} cloud services.

Before the era of \gai{} cloud services, traditional ML platforms like AzureML, AWS SageMaker, and Google Cloud ML were primarily used for tasks such as training, inference, and model fine-tuning~\cite{xin2021production}. 
These services have been well-studied for issues like deployment challenges, fault taxonomy, and bug characteristics~\cite{DBLP:conf/sigsoft/IslamNPR19, DBLP:conf/icse/HumbatovaJBR0T20, chen2020comprehensive}, while extensive research has similarly examined incident management practices, root causes, and triage procedures in conventional cloud services~\cite{chen2020towards, how_to_fight, dogga2023autoarts, bugs_make_production, chen2024automatic, how_long_to_mitigation, incident_triage_oss}. 
However, \gai{} cloud services fundamentally differ from these.
Specifically, \gai{} services such as large language models (LLMs) rely on massive parameter scales, high hardware demands, and provide natural language-driven applications like text generation, summarization, and translation, which traditional cloud services do not~\cite{min2022rethinking, liddy2001natural}.
These services also allow users to fine-tune models using user-uploaded datasets~\cite{fine-tune}, exposing risks from model-level behavior changes. Moreover, they provide intuitive conversational user interfaces, making them accessible to a broader audience while adding complexity and risks in managing user interactions. 
Such characteristics create new reliability issues related to model quality, privacy, and performance, layered atop conventional reliability concerns. 
Therefore, due to the distinctive challenges of \gai{} cloud services, it is necessary to investigate \gai{} incident patterns, impacts, and mitigation strategies to ensure future dependable and reliable GenAI services. 

In this study, we examine incidents in the \gai{} cloud service of \cloud, a leader in the GenAI field, known for hosting GPT series models. 
\cloud's Incident Management system (IcM) documents a wide range of incident data, including root causes, mitigation steps, and detailed engineer discussions, enabling a comprehensive and comparative analysis of \gai{} cloud service incidents alongside conventional cloud services. 
Our investigation reveals that while some traditional reliability challenges, such as system downtime or latency issues, remain relevant in \gai{} services, new and unique challenges have emerged. 
For example, incidents like response quality degradation show that models can unexpectedly produce low-quality or even inappropriate output from simple prompts. 
We term these incidents \textit{\gai{} incidents}.

Our study leads to crucial findings. For instance, we find that 
(1) \gai{} incidents manifest as performance degradation (49.8\%), deployment failure (35.7\%), and invalid inference (14.5\%),
significantly impacting both service reliability and user satisfaction;
(2) \gai{} cloud services experience a higher rate of incidents detected by humans (38.3\%) compared to other services (13.7\%) rather than automated monitors. Also, there is a higher false alarm rate for \gai{} (11.0\%) versus other services (3.8\%);
(3) Due to human-reported nature, many \gai{} incidents need to be re-assigned to different teams, 
and \gai{} incidents need more time (1.12 time units on average) to mitigate compared to those in other services (0.65 time units on average); 
(4) During mitigation, a specific root cause is not tied to a single type of fix. For example, while code bugs account for 21.5\% of the \gai{} incidents, only 7.6\% of fixes are code changes, with other strategies being employed. Given the tight deadlines for on-call engineers, quick approaches like rollback are prioritized to reduce downtime.

In summary, this paper makes the following main contributions:
\begin{packed_itemize}
\item We make the first attempt to unravel the general behavior of incidents occurring in \gai{} cloud services by collecting and analyzing  
a large number of \gai{}-related incidents from \cloud\footnote{Due to company policy, we hide the actual numbers and present normalized numbers in this paper.}.
\item We identify not only the symptoms and impacts of high severity \gai{} incidents but also uncover the root causes behind them and how they were mitigated with many real-world incident cases.
\item We reveal the challenges of handling \gai{} incidents at different incident life cycles and provide insights into improving the reliability of large-scale \gai{} cloud services.
\end{packed_itemize}

\section{Background and Motivation}

\begin{figure}[t]
  \centering
  \includegraphics[width=.9\linewidth]{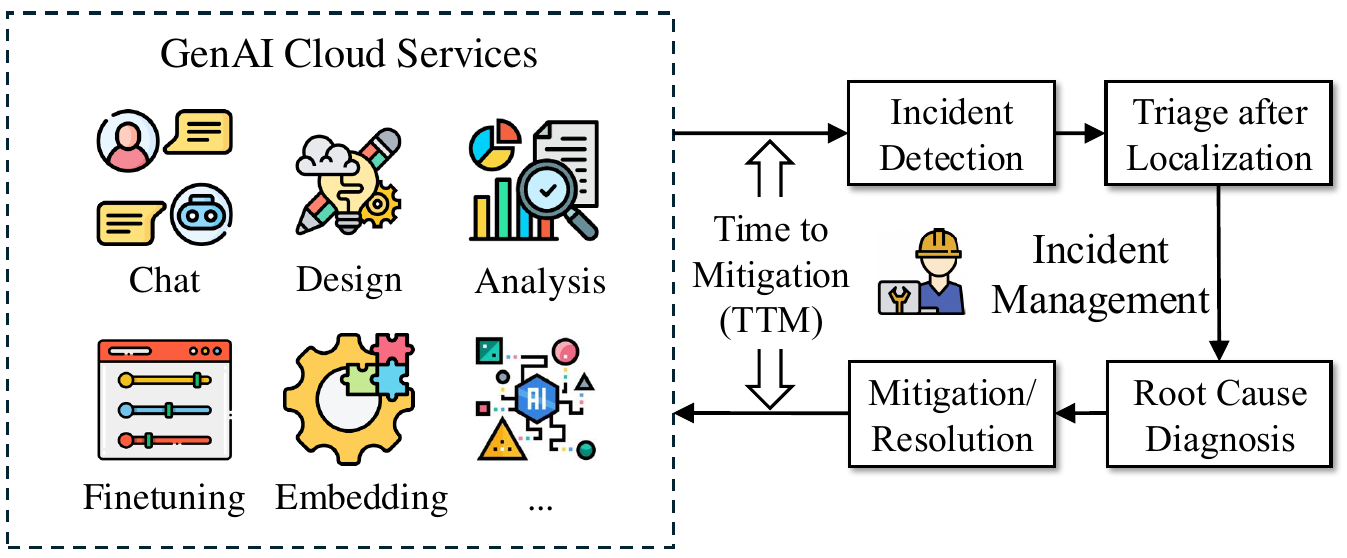}
  \caption{Incident management of \gai{} cloud services.}
  \label{fig:0}

\end{figure}

In this section, we begin by introducing Large Language Model (LLM) cloud services and incident management, as illustrated in \autoref{fig:0}. Subsequently, we outline the motivation behind our study. 

\subsection{\gai{} Cloud Service}

With the substantial parameter scale of foundation models such as GPT-4, they are typically deployed in cloud systems like Azure OpenAI. This Cloud4AI service offers users a convenient means to access advanced language models without the complexities of managing infrastructure or undertaking extensive local computations. \gai{} also has APIs for cloud services, as seen with Copilot~\cite{microsoftcopilot}, referred to as AI4Cloud. In our study, both Cloud4AI and AI4Cloud are the subjects of our investigation, which we collectively refer to as \gai{} cloud services.
In our study of \gai{} cloud services' incidents (termed as \gai{} incidents), we collect incident data from the \cloud{} Incident Management system (IcM)~\cite{chen2020aiops} (\autoref{sec:methodology}).

\subsection{Incident Management}
\label{sec:incident_management}

In cloud services, incidents are common and can lead to service disruptions, economic losses, and other unexpected severe consequences. To address such issues, major cloud providers like \cloud{} typically involve four main procedures: detection, triage, diagnosis, and mitigation (\autoref{fig:0}).

\begin{itemize}[leftmargin=*]
\item \point{Detection} This step detects service violations or performance issues and creates a ticket to record relevant information~\cite{DBLP:failure_detection,huang2023twin,zhang2022deeptralog,Warden,qiu2020firm,ma2021jump,zeng2021watson,zeng2022shadewatcher, zhang2024end, yu2024pre, yudong2024earkybird, jun2024deliver, sun2024art}. 
Such incidents can be detected manually (e.g., by customers or engineers) or automatically (e.g., by the service monitor)~\cite{DBLP:failure_detection,huang2023twin,zhang2022deeptralog,Warden,qiu2020firm,ma2021jump,zeng2021watson,zeng2022shadewatcher, zhang2024end, yu2024pre, yudong2024earkybird, jun2024deliver, sun2024art}.
\item \point{Triage} This process assigns the detected incident to a responsible team~\cite{bansal2020decaf, chen2019empirical, chen2019continuous, wang2024large, sun2024art}. 
Due to the complexity of cloud service systems, determining the appropriate team may require multiple rounds of discussions, and reassignment is also necessary.
\item \point{Diagnosis} The assigned team analyzes the incident to determine its root cause by examining system logs and configuration settings to isolate the problem and identify corresponding factors~\cite{wang2023rcagent,zhang2023pace,jin2023assess,yu2023nezha,lee2023eadro,zhang2023robust,zhang2021cloudrca, xie2024microservice, zhang2024failure, sun2024art, ding2023tracediag, zhang2024automated, ma2020diagnosing}.
\item \point{Mitigation} The mitigation step often accompanies the diagnosis, as engineers strive to promptly resolve the incident to minimize the Time to Mitigate (TTM)~\cite{jiang2020mitigate, ahmed2023recommending, zhang2024deoxys, li2024can}. 
\end{itemize}

\subsection{Motivation}
\label{subsec:motivation}

\begin{figure}[t]
  \centering
  \includegraphics[width=\linewidth]{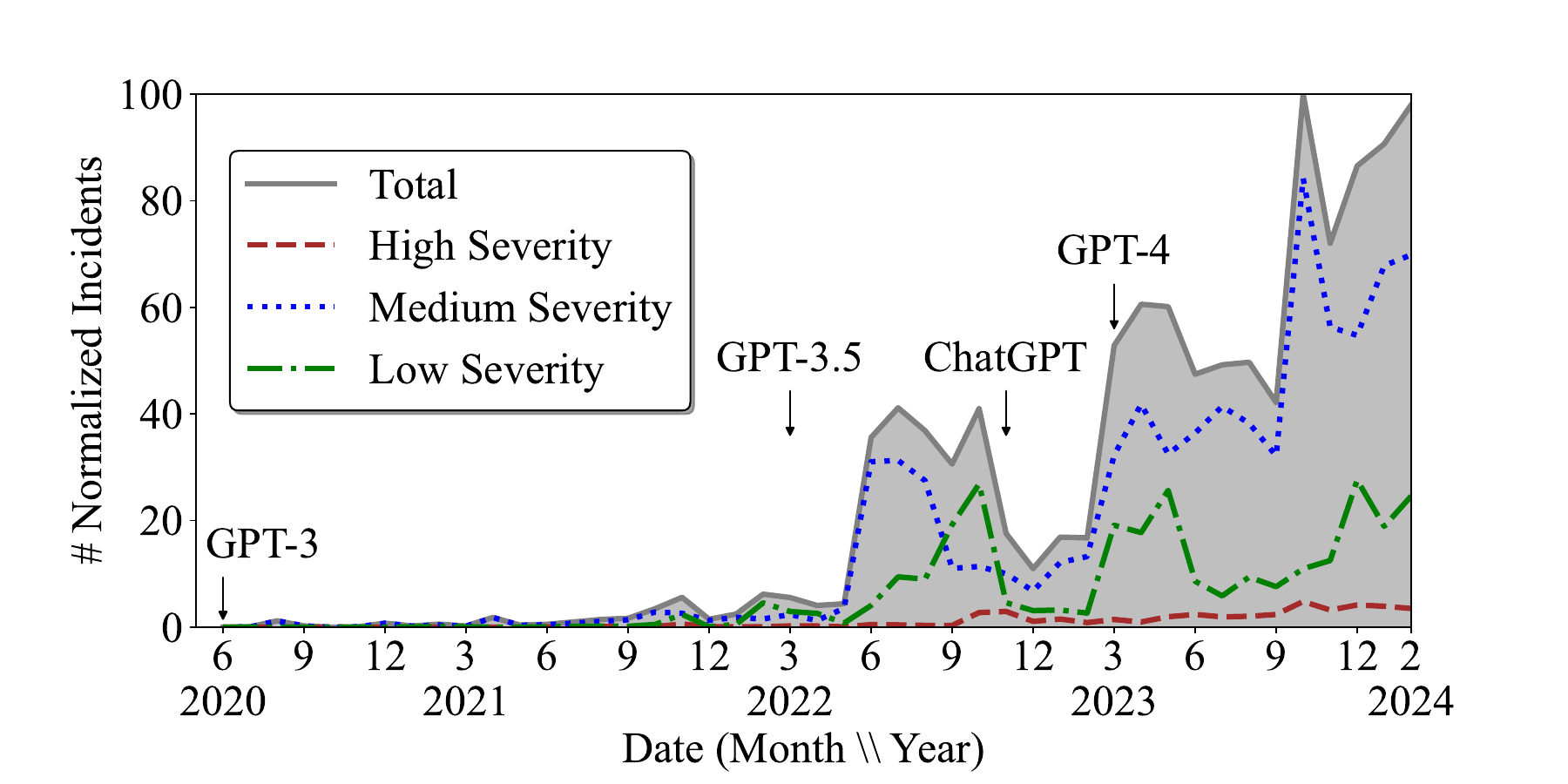}
  \caption{Number of \gai{} incidents at different time.}
  \label{fig:incident_occur}
\end{figure}

\gai{}, especially LLMs such as OpenAI's ChatGPT, has witnessed a surge in their popularity, with ChatGPT having over one million users in its debut week. However, such increased adoption has also unveiled potential risks, including outages, and errors. \autoref{fig:incident_occur} showcases the variation in terms of the number of \gai{}-related incidents within \cloud{} over the recent four years, also highlighting changes in the number of incidents across different severity levels. The lower the severity level, the higher the impact to customers. 

The total number of incidents in gray color shows an upward trend. Specifically, before the release of the GPT-3.5 model in March 2022, \gai{}-related incidents account for a mere 3\% of the total incidents within the \gai{} cloud service. 
After 2023, there is a significant increase in incidents, with a pronounced spike following the introduction of GPT-4 in March 2023. At this point, the volume of incidents had increased nearly tenfold relative to the figures reported during the GPT-3.5 era. This dramatic rise can be attributed to the global fame attained by the GPT model, which attracted millions of users.
This trend also holds across all severity levels, with lower-impact incidents comprising most cases.

The proliferation of \gai{}-related incidents affects both the associated cloud services and their end users.
Unfortunately, the characteristics of the incidents of \gai{} cloud services have not yet been comprehensively unveiled. 
This study aims to bridge this gap, thus providing insights for future research and practical guidance for the software engineering community maintaining \gai{} cloud services.

\section{Methodology}
\label{sec:methodology}

\cloud, a leader in cloud computing, hosts the training and APIs for OpenAI and offers various \gai{} cloud services, including \aoai, which utilizes \cloud~platform to provide access to the GPT series models. Incidents in these services are documented in a dedicated database. Prior researches~\cite{bugs_make_production, changed_incidents, how_long_to_mitigation, how_to_fight} have utilized similar database to collect incidents and derive analytical insights. Consistent with this approach, this study leverages \cloud's database to collect \gai{}-related incidents.

The database contains key details for each incident, including its description, root cause, mitigation steps, discussions by the on-call engineers (OCEs), and severity-level tags (high, medium, and low). To conduct our empirical study, we collect both \gai{}-related and non-\gai{} incidents as a comparative dataset. Following the methodology of previous research~\cite{how_to_fight}, we focus on significant incidents characterized by their high severity and detailed root cause descriptions, thereby facilitating an insightful qualitative analysis.
The following shows the details of our methodology.

\subsection{Data Collection}
We first introduce the detailed procedures to collect the dataset.
In particular, we collect two datasets that serve for the 
two incident study respectively. 
The general incident study is designed to explore general characteristics of \gai{} incidents within the incident management process, such as the distribution of incidents' detection methods. It also aims to compare these incidents with those from other cloud services, analyzing the differences between the two.
It requires a dataset with broad coverage. Therefore, we endeavor to collect all incidents that meet the criteria as comprehensively as possible. 
The in-depth incident study focuses on understanding the categories of an incident's symptoms, root cause, and mitigation strategies based on detailed information, such as discussions by OCEs. 
Given the large volume of data, we opt to select only high-severity incidents as in-depth analysis cases, as these incidents have a more significant impact on the system and tend to attract greater attention. 
Through both the study, we can comprehensively understand the characteristics of \gai{} incidents. 

\point{$\bullet$~\gai{} incidents collection for general analysis} 
In this phase, our primary goal is to gather data on incidents, 
encompassing the period from June 2020, following the release date of GPT-3 model by OpenAI, to February 2024.
Here are the criteria we use to collect \gai{}-related incidents: 
\begin{enumerate}[leftmargin=*]
\item We choose incidents that have been mitigated or resolved. The incident status is categorized within the ``Status'' field as ``Active'', ``Mitigated'', and ``Resolved''. Our collection excludes incidents marked as ``Active'' due to the lack of comprehensive data, such as discussions by OCEs, root cause analysis, and mitigation steps; 
\item The ``Service'' field indicates whether the incident is associated with a \gai{} cloud service and its team or not. Incidents are considered \gai{}-related if they are linked to a specific \gai{} cloud service, such as \aoai;
\item  Given the complex architecture and dependencies of \gai{} cloud services, certain \gai{} incidents may be managed by dependent (sub-)services and cannot be directly found by ``Service''. Thus, we define a vocabulary of words related to
\gai{} (\textit{e.g.}, ``gpt-3.5-turbo'', ``LLM'', \textit{et.al}). 
Then we perform a case-insensitive search of these terms within the
``Title'' of an incident. 
\end{enumerate}

Following these criteria, we obtain hundreds of thousands of %
\gai{}-related incidents.

\point{$\bullet$~\gai{} incidents collection for in-depth analysis}
We meticulously select a subset of \gai{} incidents based on three criteria: 
\begin{enumerate}[leftmargin=*]
\item The incident must be of high severity. Incidents of this nature typically result in significant service disruptions, affecting numerous tenants and customers; 
\item The incident should include a detailed root cause analysis; 
\item The incident must be valid. We deem an incident as invalid if its mitigation steps are described as a ``False Alarm''. 
\end{enumerate}

Following these criteria, we identified and selected many
incidents for our detailed analysis. 
Given that high-severity incidents inherently constitute a smaller proportion of total incidents, the data collected at this step is significantly less than what is gathered for qualitative analysis. 

\point{$\bullet$~Other incidents collection}
For discussion, especially a comparative mitigation analysis between \gai{} incidents and those unrelated to \gai{}, we collect %
the same number of other incidents using the same time frame and criteria in general analysis (omitting the (2) and (3)) and in-depth analysis. 

\subsection{Research Questions}

In this study, we aim to reveal the behaviors of \gai{} incidents in the incident management life cycle.
Such insights are critical for the development, maintenance, and management of LLMs, aiming to improve the robustness and reliability of the LLM cloud systems. 
This exploration is pivotal for providing a scientific basis to prevent future incidents, thereby contributing valuable knowledge and experience to both the research and practical applications in the field.
In particular, we design the following research questions (RQs). 

\point{RQ1} What is the general behavior of \gai{} incidents in terms of different incident life cycles? 

\point{RQ2} What are the symptoms of \gai{} incidents?

\point{RQ3} What are the root causes of \gai{} incidents?

\point{RQ4} How are \gai{} incidents mitigated?

\subsection{Categorization Strategy}
\label{sec:Categorization Strategy}

While each incident is documented with detailed information, 
these records are typically composed by humans, e.g., OCEs' discussion, and may contain images, URLs, and other elements that complicate automatic categorization. Therefore, we need to analyze all the incidents manually to further understand their symptoms, root causes, and mitigation strategies. 

We divide our dataset of incidents into three subsets randomly: (1) \textit{taxonomy set}: 40\% incidents, (2) \textit{validation set}: 20\% incidents, and (3) \textit{test set}: 40\% incidents.
Firstly, two authors independently following the open coding strategy~\cite{strauss1997grounded} to label both symptoms, root causes and mitigation strategies for the \textit{taxonomy set}. 
Next, for categories with inconsistent classifications, a meeting involving other authors will be convened to determine the final categorization. 
The two authors then label the \textit{validation set} to check for the emergence of new categories to perform further discussions to refine their understanding of each category. 
Finally, they label the \textit{test set} and employ Cohen's kappa~\cite{kappa} coefficient to measure the consistency between annotators. 

After multiple rounds of the labeling process described above, we ultimately adopt the best result, achieving near-perfect agreement across the three taxonomies: Symptom: 0.921, Root Cause: 0.930, Mitigation: 0.893. 
For incidents that can fit into multiple categories, e.g., multiple symptoms, disagreements are resolved by focusing on the category most prominently reflected in the incident and OCE's discussions.

\section{RQ1: General Statisitcs}
\label{sec:gs}
We explore the characteristics of \gai{} incidents from three aspects, \textit{detection}, \textit{triage}, and \textit{mitigation}, each corresponding to a phase of the incident life-cycle.

\subsection{Incident Detection}
\label{subsec:llm_detection}
\begin{figure}[t]
    \centering
    \begin{tcolorbox}[colback=white, colframe=white, width=\columnwidth, boxrule=.1pt]
    \ttfamily
    \textbf{Title:} monitor evaluated high fail rate for scope [ServiceA], zone [WestRegion2] \\
    \textbf{Monitor Name:} 
    [Service]\_FailRate\\
    \textbf{Metric:} DependencyCallCounter\\
    \textbf{Description:} Marks the target as `Unhealthy' and raises a high-severity incident if the failure rate exceeds 4\% over the past 60 minutes. \\
    \textbf{Trouble-shooting Guide:} [Link to the TSG]

    \textbf{Diagnostic Information:} \\
    \begin{tabular}{|p{0.8\columnwidth}|r|}
    \hline
    \textbf{Failure Description} & \textbf{Count} \\
    \hline
    ServiceClient failure for [ServiceB]: Failed to call [ServiceB], ReasonPhrase=Failed Dependency & 52 \\
    \hline
    RequestTimeout for [EncoderService] & 13 \\
    \hline
    ServiceClient failure for ChatGPT: No service for `BotClientLibrary' has been registered. & 8 \\
    \hline
    ServiceClient failure for ChatGPT: Failed to call `ChatGPT' at LoadBalancer, ErrorStatusCode=400 & 3 \\
    \hline
    \textbf{ClientSecretCredential authentication failed: A configuration issue is preventing authentication. Details: The provided client secret keys for app [ApplicationA-UUID] are expired.} & 6 \\
    \hline
    ... ... & \\
    \hline
    \end{tabular}
    \end{tcolorbox}
    \caption{Incident detected by a monitor and the collected diagnostic information attached to the monitor.}
    \label{fig:monitor}
\end{figure}

\begin{figure}[t]
\centering
    \begin{subfigure}[b]{0.22\textwidth}
        \centering
        \includegraphics[width=1\linewidth]{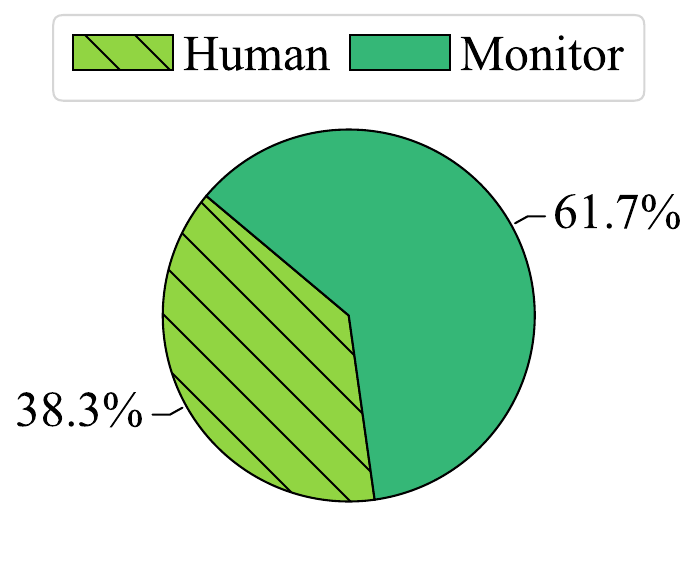}
        \caption{Detection type.}
    \end{subfigure}
    \begin{subfigure}[b]{0.24\textwidth}
         \centering
         \includegraphics[width=1\linewidth]{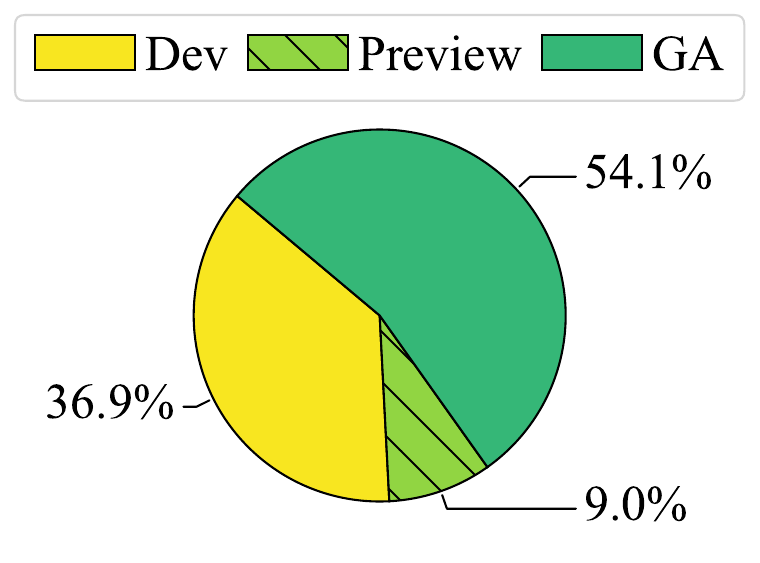}
         \caption{Service stage.}
     \end{subfigure}
\caption{\gai{} incident detection type and different stages of \gai{} services. GA: General Availability, Dev: Development.}
\label{fig:detection}
\end{figure}

Detection is the initial step in incident management for a cloud service. Engineers can identify incidents by noticing unusual system behaviors~\cite{incident_triage_oss, DBLP:conf/www/time_series_detection, DBLP:journals/pvldb/ImDiffusion, DBLP:conf/icse/TraceArk}, while customers can also report issue tickets when encountering failure messages or experiencing delay ~\cite{DBLP:conf/sigsoft/customer_report}. To improve the efficiency of incident detection, automated monitoring tools are deployed~\cite{DBLP:failure_detection}. These tools either passively collect real-time system telemetry data (e.g., CPU usage) and performance measures (e.g., response time and throughput), or proactively check the health of the system by periodically performing heartbeats or sanity checks. \autoref{fig:monitor} shows a monitor detecting the calling failure rate of a service.%

\noindent\textbf{Missing Alarms (False Negative):}
As shown in \autoref{fig:detection}, 
we find that 38.3\% of the incidents related to \gai{} are reported by humans, such as engineers and customers, instead of automated incident monitors. 
To explain this high human-reported percentage (\ie~such a ratio is only 13.7\% for other cloud services, as will be further discussed in \autoref{subsec:lessons}), we find that 45.9\% of \gai{} cloud services are still under development or in the preview stage, while 54.1\% of \gai{} cloud services are in the General Availability status.
Moreover, many \gai{} cloud service monitors currently build on adaptations of existing frameworks designed for other types of cloud services,
which may not yet fully align with the unique requirements of \gai{}-specific scenarios. 
For instance, invalid inference incidents are often identified and reported by users, reflecting the collaborative effort to refine these systems further.
Our study observes that there are around 25.9 unique monitors per 100 monitor-reported GenAI incidents, compared to 74.4 for other cloud services, offering an opportunity to enhance monitoring diversity.
These insights highlight the ongoing evolution of GenAI monitoring approaches, as the industry continues to refine automated detection capabilities and improve response efficiency.

\begin{table}[t]
  \centering
  \caption{Detection type distribution and false alarms rate for \gai{} and non-\gai{} incidents.}
  \label{table:false_alarm_and_detection}
  \begin{tabular}{lcccc}
    \toprule
    & \multicolumn{2}{c}{\textbf{Detection Type}} & \multicolumn{2}{c}{\textbf{False Alarm Rate}} \\
    \cmidrule(lr){2-3} \cmidrule(lr){4-5}
    & Human & Monitor & Human & Monitor \\
    \midrule
    \textbf{\gai{}}  & 38.3\% & 61.7\% & 6.6\% & 11.0\%\\
    \textbf{Other} & 13.7\% & 86.3\% & 4.8\% & 3.8\% \\
    \bottomrule
  \end{tabular}
\end{table}

\noindent\textbf{Wrong Alarms (False Positives):}
The false alarm rate for incidents detected by monitors in \gai{} services is notably high at 11.0\% (\autoref{table:false_alarm_and_detection}),
compared to the 6.6\% detected by humans. 
This higher false positive rate is primarily from the sensitivity of the monitoring systems. 
For example, the monitor in \autoref{fig:monitor} issues an incident report if the failure rate exceeds 4\% within one hour. 
If the failure rate threshold is set lower or the monitoring period is shortened, the monitor becomes more sensitive, possibly leading to more false positives.
These false alarms burden engineers 
with unnecessary investigations, thus delaying the resolution of true incidents.

\insight{
\gai{} cloud services and their monitoring are still in an early stage. A high percentage (38.3\%) of \gai{} incidents are reported by humans. Besides, among the incidents detected by the automated monitors, there is an 11.0\% false alarm rate, which points to opportunities for further enhancement in monitoring precision.
\label{Finding:1}
}

\subsection{Incident Triage}
\label{subsec:LLM_triage}

Triage is a crucial component of the incident management life cycle, significantly affecting the Time-to-Mitigate (TTM)~\cite{incident_triage_oss}. Incidents can be sent to incorrect teams or need collaborative efforts, leading to cases where they are re-assigned between different teams. The process of reassigning an incident from one team to another is called a \textit{transfer hop}.

\begin{figure}[t]
  \centering
  \includegraphics[width=0.9\linewidth]{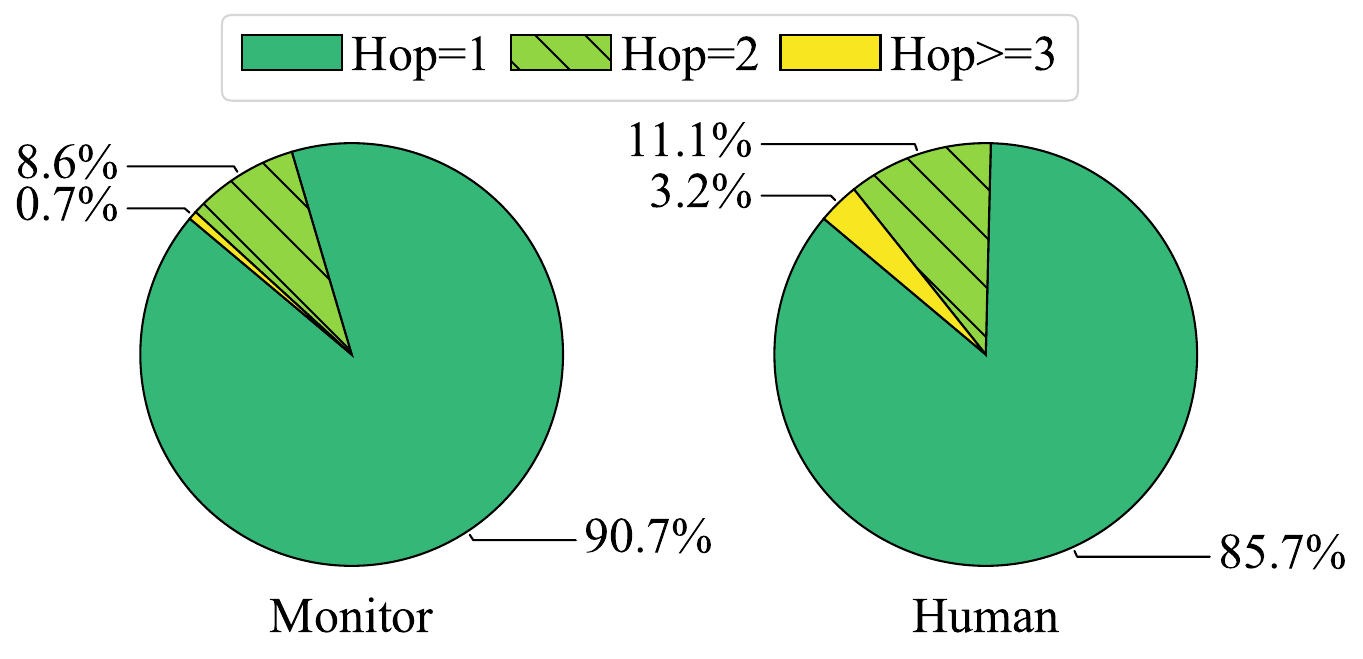}
  \caption{Transfer hops for incidents.}
  \label{fig:1}
  \label{fig:triage}
\end{figure}

As shown in \autoref{fig:triage}, incidents that are initially detected by monitoring systems are usually accurately triaged to the correct team on their first attempt (90.7\%).
However, the proportion of incidents needing triage increases when detected by humans. 
\gai{} incidents detected by humans that undergo reassignment is 14.3\%. 
This shows the effectiveness of using automatic monitors for triage. For example, the monitor-generated ticket title embeds the name of the service that leads to the incident, as shown in \autoref{fig:monitor}, so the incident can be accurately triaged to the service team. 
Another factor for the incident re-assignment is the interdependency on other services. 
Resolving an incident might exceed the capabilities of a single team, and collaborative efforts across different service domains are needed.
Further details on the root causes of \gai{} incidents will be elaborated in \autoref{section:rc}.

\subsection{Incident Mitigation}

\begin{figure}[t]
\centering
    \begin{subfigure}[b]{0.22\textwidth}
        \centering
        \includegraphics[width=1\linewidth]{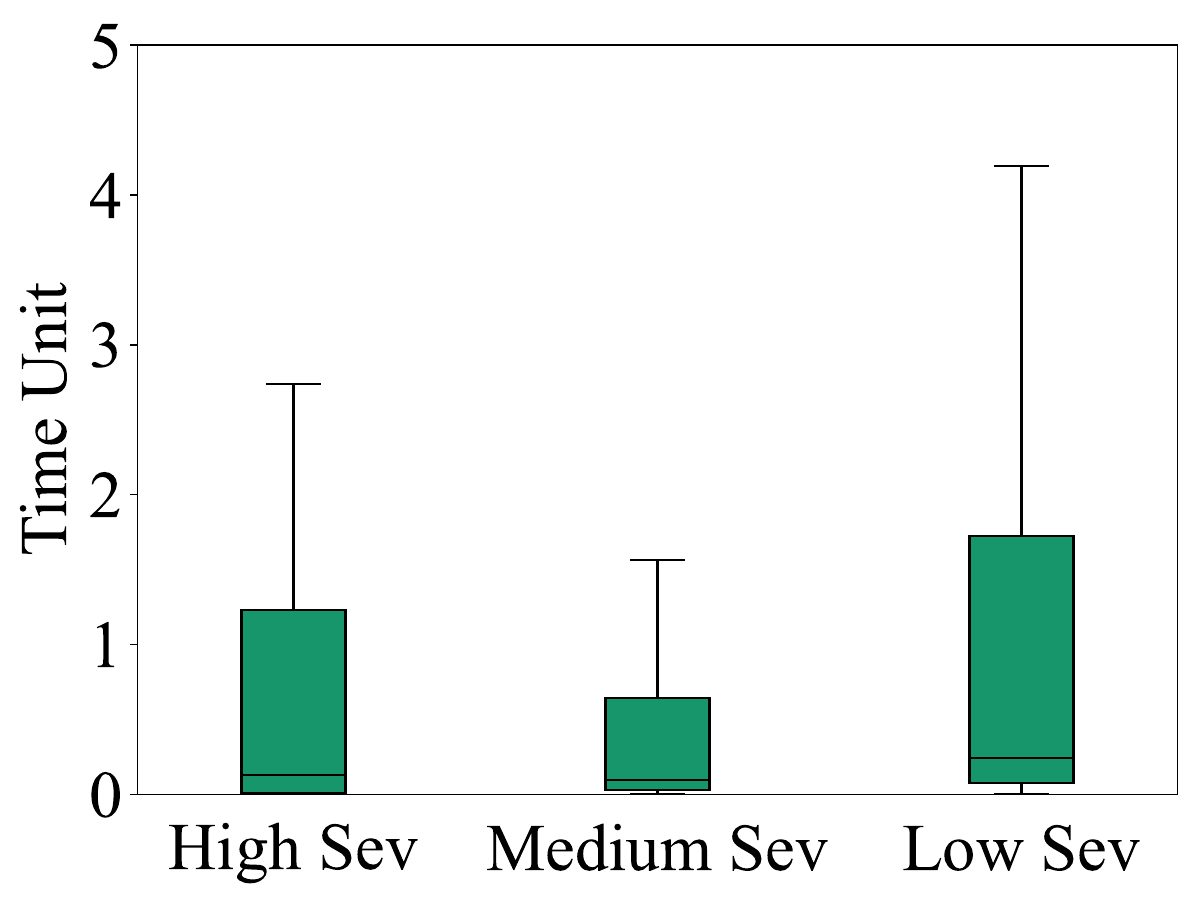}
        \caption{}
        \label{fig:ttm_all_a}
    \end{subfigure}
    \begin{subfigure}[b]{0.11\textwidth}
        \centering
        \includegraphics[width=1\linewidth]{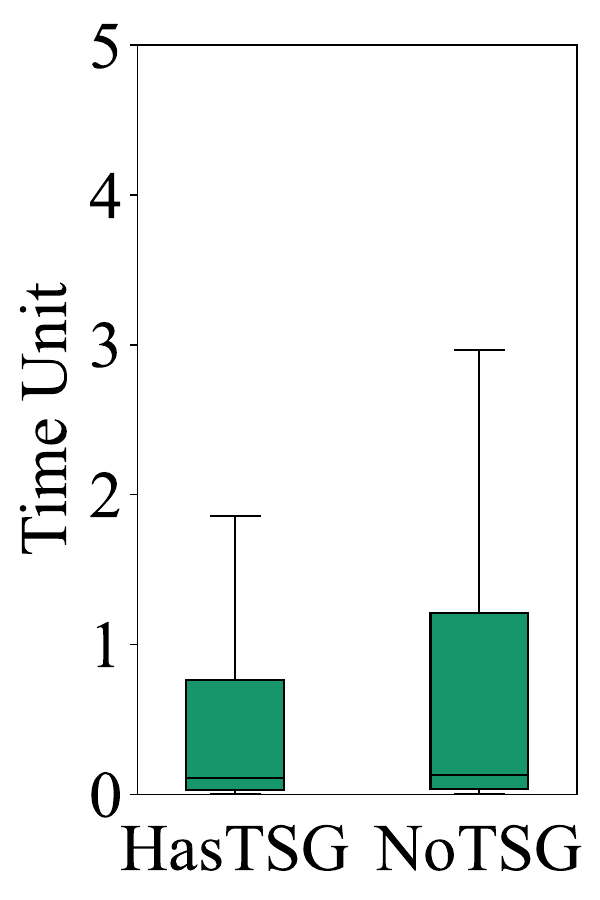}
        \caption{}
        \label{fig:ttm_all_b}
    \end{subfigure}
    \begin{subfigure}[b]{0.11\textwidth}
        \centering
        \includegraphics[width=1\linewidth]{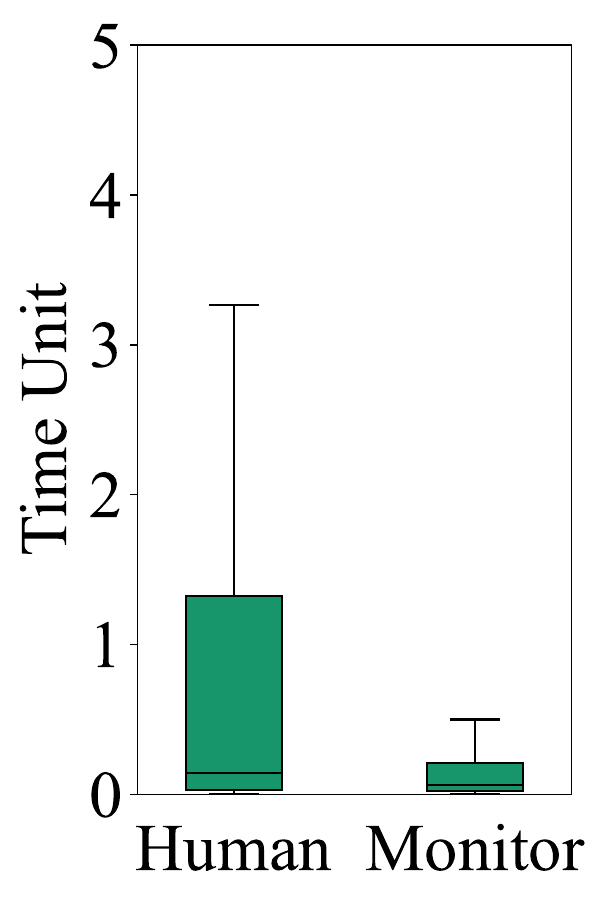}
        \caption{}
        \label{fig:ttm_all_c}
    \end{subfigure}
\caption{TTM distribution across different factors: Y-axis is the normalized TTM of all incidents; the top whisker of each box plot represents the maximum value; the top and bottom edge of the box represent the upper quartile and the lower quartile, respectively, and the line inside the box represents the median value. (a) Different severity levels; (b) The presence of a TSG; (c) Detection types.}%

\label{figure:ttm_all}
\end{figure}

Intuitively, we would expect incidents with higher severity to have longer TTM, as these incidents usually require more extensive investigation and resolution efforts. For example, as shown in \autoref{fig:ttm_all_a}, high-severity incidents generally take longer to resolve than medium-severity.
However, low-severity incidents exhibit a significantly longer TTM compared to other severity levels because these lower-priority \gai{} incidents often remain unresolved for extended periods due to their low impact.

\label{subsec:LLM_mitigate}

Incidents accompanied by a TSG are resolved more swiftly than those without one, as TSGs provide clear guidelines and solutions that facilitate faster mitigation (\autoref{fig:ttm_all_b}). Furthermore, our analysis reveals that incidents generated by monitors are mitigated more quickly than those reported by humans (\autoref{fig:ttm_all_c}). This is partly because monitors, as illustrated in \autoref{fig:monitor}, often include links to corresponding TSGs. 
By following the TSG instructions, diagnostic information is more readily collected. For example, it becomes immediately clear that the root cause of the incident in \autoref{fig:monitor} is expired secret keys of the application, thereby enabling quicker resolution.

\insight{Automatic monitors and trouble-shooting guides (TSGs) can significantly boost the mitigation process, and reduce the Time to Mitigation for \gai{} incidents. 
}
\label{finding-ttm}

\section{RQ2: Symptom of \gai{} Incidents}
\label{section:symptom}

We analyze the symptoms of \gai{} incidents from available incident-related telemetry data (metrics, logs, traces, etc.) and discussion threads from on-call engineers. We categorize the symptoms into invalid inference, deployment failure, and degraded performance. Note that one incident may have multiple symptoms, and we choose the major symptom as its category as mentioned in \autoref{sec:Categorization Strategy}.
The following subsections are ordered based on their perceived impact on service operation and user experience.

\subsection{Invalid Inference (14.5\%)} 
While the model inference executes successfully and the service returns results to clients without errors, the model output can be invalid. 
Inaccuracies in the output directly affect the core functionality of \gai{} services.
\textit{(1) Response Quality Degradation (10.7\%):}
Models can generate low quality content with even simple user prompt. 
Another scenario involves the generation of invalid content, where the model could not understand the user's prompt, leading to invalid content creation \cite{wester2024ai};
\textit{(2) Prompt/Response Content Filter Malfunction (3.8\%):} \label{point:contentfilter}
\gai{} cloud services deploy policy filters for both user prompts and model responses to prevent the generation of harmful content. However, these content filters can sometimes malfunction, resulting in inappropriate or harmful content from the model, as well as false alarms that incorrectly filter out valid prompts or responses. %

\subsection{Deployment Failure (35.7\%)}

Deployment failures reflect the impact on \gai{} service continuity.
We find: 
(1) \textit{Model deployment failure (12.0\%)}: 
When users are training or fine-tuning large language models, the deployment failure may happen. For instance, all of the user fine-tuned models were not successfully deployed in time for a specific deployment region;
(2) \textit{Resource deployment failure (14.4\%)}:
\gai{} cloud services heavily depend on different types of resource deployment, like computing, networking and storage resources for consuming, transmitting and storing vast volumes of data.
Failed deployment of these can propagate exceptions to other parts of the \gai{} services;
(3) \textit{Fine-tune API failure (9.3\%)}: 
\gai{} cloud services offer interfaces for uploading/downloading data, model selection, and parameter setting, which users can customize to fine-tune their own models. However, a failure may happen when calling such fine-tune REST APIs. For instance, a conflict version requirement caused the failure of the fine-tune API calls.

\vspace{1pt}
\subsection{Degraded Performance (49.8\%)}
\vspace{1pt}
There are two typical performance degradation:
(1) \textit{Service-level Degradation (27.2\%)}: Multiple APIs within a \gai{} service can fail simultaneously, impacting the overall availability and performance of that service. 
Also, if multiple service nodes become unhealthy, e.g., out-of-memory or disk pressure, the performance of the whole service can be influenced;
(2) \textit{API-level Degradation (22.6\%)}: A particular \gai{} API can be 
delayed.
Degraded performance is primarily due to infrastructure and configuration issue as discussed in \autoref{section:rc}.

\insight{
\gai{} incidents can occasionally include challanges, 
including invalid inference (14.5\%), deployment failure (35.7\%), and performance degradation (49.8\%). 
\label{finding:symptoms}
} 

\vspace{5pt}
\section{RQ3: Root Cause}
\label{section:rc}

\begin{figure}[t]
  \centering
   \includegraphics[width=.9\linewidth]{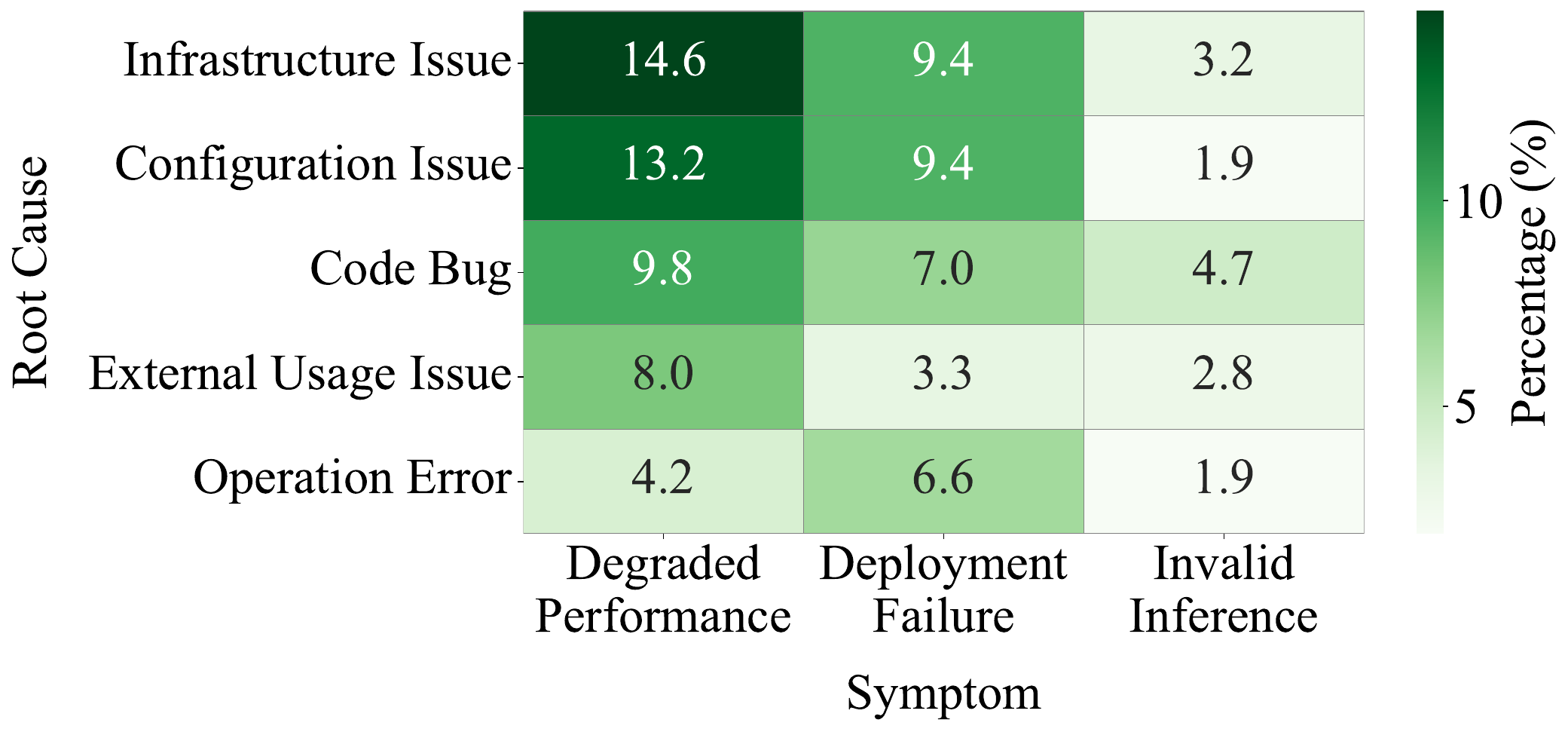}
  \caption{Relationships between symptom and root cause.}
   \label{fig:symptom}
\end{figure}

We categorize the root causes of \gai{} incidents into five distinct types.
The relationships between symptoms and root causes are shown in ~\autoref{fig:symptom}.
Each cell represents the percentage of a specific symptom associated with a particular root cause.
We can observe that a single symptom can come from multiple root causes rather than a simple one-to-one relationship. This indicates that diagnosing the root cause from symptoms is not straightforward.

\subsection{Infrastructure Issue (27.2\%)} 
\label{subsec:infra}

\gai{} cloud services are built upon a complex hierarchical infrastructure comprising VMs, nodes, clusters, and data centers that host tightly coupled resources, including CPU, memory, storage, and networks. 
We find that infrastructure issues are a major cause of degraded performance and deployment failure (\autoref{fig:symptom}). The infrastructure is categorized into the following types: 
(1) \textit{Infrastructure Maintenance Issues (17.8\%)}: Failures of hardware components, such as worn-out GPUs, can impact the fine-tuning and inference of \gai{} services. For instance, faulty GPUs can process requests incorrectly, resulting in errors such as gibberish outputs.
(2) \textit{Network Issues (4.7\%)}: Besides the network bandwidth, incidents can happen between the communication of VMs and nodes within clusters, including connectivity issues and DNS resolution failures. Such network problems can severely disrupt the performance and reliability of the service. %
(3) \textit{Storage Issues (4.7\%)}: The management of vast amounts of data needs robust storage solutions. Failures in data storage or IO operations, such as data corruption or delays, can lead to service disruptions. 

\vspace{5pt}

\insight{
Infrastructure issues are a key area of focus for understanding and addressing incidents in \gai{} cloud services, especially for degraded performance and deloyment failure.
To meet the growing user demands, \gai{} cloud services should not only scale up the size of GPU cluster but also prioritize robust infrastructure management.
}\label{finding:7}
\subsection{Configuration Issue (24.5\%)}

\gai{} cloud services rely on a multitude of configuration settings to ensure the seamless operation of their interconnected components. However, mismanagement of these configurations is occasionally observed. Incorrect or unsynchronized settings can ruin service functionality.
We categorize these configuration issues into the following types:
\textit{(1) Misconfiguration (13.1\%)}: Operators may employ incorrect configurations or commit errors, typically due to human mistakes. For example, engineers might configure much fewer model instances than required during system maintenance, leading to an outage of degraded performance. (2) \textit{Configuration Update (6.4\%)}: Changes in one cloud component's configurations can lead to incompatibilities with other components due to the configuration dependencies among them. Additionally, version conflicts for the same configuration may result in one configuration overriding another, e.g., using a removed parameter in its latest version or using an added parameter in its previous version, leading to malfunctions. (3) \textit{Configuration Missing and Gaps (5.0\%)}: Missing or disabled configurations can disrupt normal operations. Additionally, certain configurations impose range restrictions on values, such as timeout thresholds or maximum sizes for prompt tokens. Under unexpected circumstances, such as a sudden surge in user traffic, these static configurations can constrain system performance.

\subsection{Code Bug (21.5\%)} Code bugs are a primary cause of incidents, and a prior work~\cite{bugs_make_production} has specifically investigated the code bugs leading to cloud incidents. 
The following shows four types of code bugs for \gai{} incidents: data constraints bugs, content filter bugs, exception handling bugs, and cross-system bugs.
(1) \textit{Bugs violating Data Constraints of the Model (6.7\%)}:
Bugs can arise due to inadequate validation for data format or missing data that the model needs to consume.
Take a fine-tune failure as an example, it can be caused by the lack of validation on dataset format in \texttt{\small FileUpload} API.
The malformed dataset was not rejected during the file upload stage, and was delivered to the backend services;
(2) \textit{Prompt/Response Content Filter Bugs (2.2\%)}: Code defects can exist in the prompt or response filter.
(3) \textit{Exception Handling Bugs (6.3\%)}: Exceptions are a normal occurrence during code execution. 
However, the code can be unable to effectively handle certain exceptions or failures.
For example, 
errors may occur during model deployment, such as an invalid model being deployed to an endpoint. Due to a code defect in processing such an error, e.g., simply swallowing the exception, the invalid model remains there and serve requests;
(4) \textit{Cross-system Bugs (6.3\%)}: These bugs are mostly caused by issues in the code across multiple components. To fix this type of bugs, changes are needed for multiple services.

\vspace{-7pt}
\subsection{External Usage Issue (14.1\%)}
Incidents can arise from incorrect usage of \gai{} service by the customer. 
For example, a customer missed indexes when performing queries to LLM, which caused the high CPU usage in the service.

\subsection{Operation Error (12.7\%)} 

Operation errors in \gai{} cloud services are typically caused by human errors during the management and operational processes. This error occurs when operators mistakenly introduce erroneous or outdated dependencies, or use expired credentials.

\section{RQ4: Mitigation}
\label{sec:mitigation}

To answer RQ4, we delve into the common categories of mitigation strategies utilized to address \gai{} incidents. 
Specifically, we inspect the title and the detailed description of the mitigation steps in each incident ticket and its corresponding postmortem report. 
these descriptions, engineers' discussion thread, and completed work bullets,
we classify the mitigation methods into the following distinct types: ad-hoc fix, self-recover, rollback, configuration fix, infrastructure fix, external fix, code fix, and others.

\subsection{Code Fix (7.6\%)} 

\vspace{1mm}
\begin{lstlisting}
  const getCommandText = () => 
      featureFlags.enableRemoveUnicodeFromRequest 
      ? removeUnicodeFromRequest(text) : text;
  ...
  
  const removeUnicodeFromRequest = (msg: string) => {
      const unescapedMsg = unescapeUnicode(msg);
      const regex = /[\u{E0000}-\u{E007F}]/gu;
      return unescapedMsg.replace(regex, "");
  };
\end{lstlisting}

This category is to address incidents by updating and fixing buggy code or by incorporating new code~\cite{bugs_make_production}, such as adding exception handling mechanisms to improve resilience or implementing new features for specific purpose.
For example, certain Unicode characters cannot be rendered in a font and thus do not appear in the user interface, resulting in what is called hidden text.
However, the hidden text can still be understood and processed by the LLM. This could be potentially exploited as an attack surface to change the response from the user's intent. 
The following code update adds a new feature to remove the Unicode characters (within the range of U+E0000 to U+E007F) that can be used as hidden text from the user's request.

\insight{Given the tight deadlines for mitigating \gai{} incidents, only a small proportion (7.6\%) of incidents are resolved through code fixes. This approach is time-consuming, requiring more efforts to design and implement the solution and navigate through an end-to-end CI/CD pipeline. Consequently, other mitigation strategies are preferred by engineers for their faster resolution times in the initial stages of mitigation.
\label{finding:code_fix}}

\subsection{Rollback (15.2\%)} 
\label{subsec:rollback}
For incidents triggered by changes, such as configuration adjustments or code updates, rollback is a widely used and efficient mitigation strategy. Engineers revert these changes to a previous, stable version. Our study identifies: 
(1) \textit{Deployment Rollback (8.9\%)}: Updates to code or third-party libraries can introduce bugs. These incidents can be addressed by reverting to a previous commit or an older stable build version of the third-party library. For example, an inference API error which caused by the compatibility issue between fine-tuning code and inference code can be fixed by rolling back to a previous inference engine for users in specific regions;
(2) \textit{Configuration Rollback (6.3\%)}: This involves undoing bad configuration changes to alleviate the issue.

\subsection{Configuration Fix (13.0\%)} 
\label{subsec:config_fix}
To address the majority of configuration errors, engineers often fix bugs in configuration files to reinstate the service. We identify two primary approaches to configuration fixes: %
(1) \textit{Add or Disable Features (7.6\%)}: Incidents can be mitigated either by adding new features that enhance service stability or by disabling features that are causing failures, thus aiding in the swift resolution of the issue; (2) \textit{Increase the Configuration Limit (5.4\%)}: Besides the configuration issues, a number of incidents from resource capacity as mentioned in \autoref{subsec:infra} can also be mitigated by configuration changes as a short-term strategy, such as increasing timeout thresholds.

\subsection{Infrastructure Fix (12.1\%)} 
\label{subsec:infra_fix}
For incidents caused by infrastructure issues, an infrastructure fix is a frequently utilized mitigation method. 
Common infrastructure fixes include scaling operations, component restarts or rebuilds, and traffic failovers.
One of the following actions can be performed: (1) \textit{Scaling~(6.3\%)}: Due to infrastructure limitations, a service may not be able to handle a large volume of traffic, and simple configuration of increasing the capacity does not work. Therefore, scaling out more instances or nodes to increase capacity is needed. For example, increasing the compute capacity allows the service to process more requests, thus avoiding an excessive number of request failures; (2) \textit{Restart or Rebuild (3.1\%)}: This category involves mitigating incidents by restarting or rebuilding faulty components; (3) \textit{Traffic Failover (2.7\%)}: This involves failing over traffic to another healthy service component, including nodes, clusters, or another cloud region. 

\insight{In practice, increasing the limit of resource configuration (5.4\%) is a straightforward mitigation strategy. However, when these configuration changes are insufficient due to the allocated resources or infrastructure reaching their capacity, re-scaling (6.3\%) becomes necessary to resolve the \gai{} incidents, even though it may take a long time to deploy the additional infrastructure.
\label{finding:infra_fix}
}

\subsection{Ad-hoc Fix (22.4\%)} 
\begin{figure}[t]
  \centering
   \includegraphics[width=\linewidth]{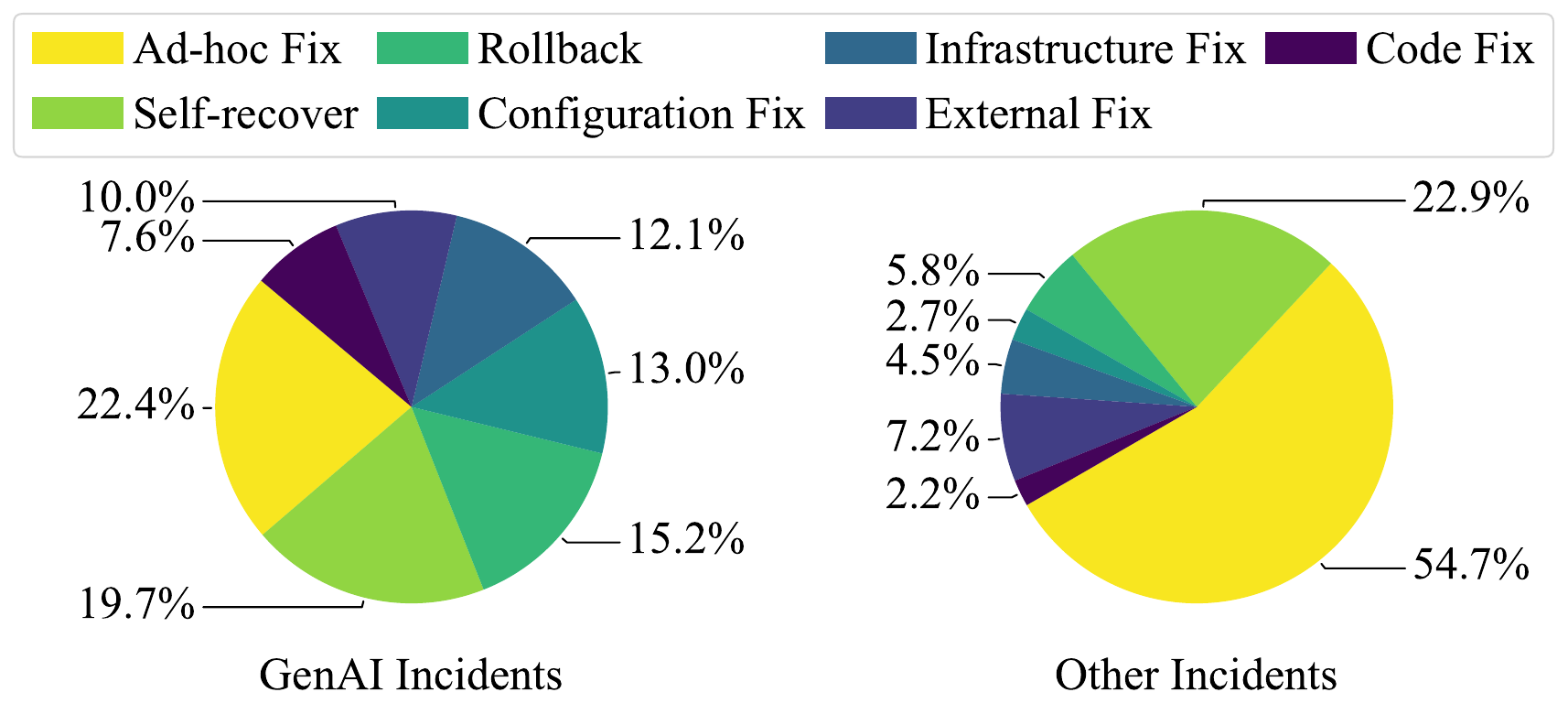}
  \caption{The distribution of mitigation approaches. 
  }
   \label{fig:mitigation}
\end{figure}

LLM incidents can be complext, and engineers may not always be familiar with the root cause of \gai{} incidents.
To address the impact quickly, standardized procedures can be costly, so a series of improvised, situation-specific steps are applied to mitigate the symptom first. 
{For instance, in response to a malicious user bypassing the batch size limitation, engineers mitigated it by identifying and blocking the malicious user, enabling the validation logic to check the \texttt{\small ImageModelA-batch-size} parameter in the request headers, and enforcing a maximum limit for the batch size.} {Also, in other cases where a single user's request consumed too many background resources and resulted in service overload, the issue was mitigated by temporarily limiting the user's request rate, adjusting the throttling from 10 seconds throttling to one second for the customer with a high workload.} Note that over half of the incidents from other cloud services are mitigated by ad-hoc fix (\autoref{fig:mitigation}), while \gai{} cloud services often require more development and deployment efforts (other mitigation approaches to be discussed in the following) to fully resolve the incidents. Consequently, the TTM of \gai{} incidents are longer.

\subsection{External Fix (10.0\%)} \gai{} cloud services support external company partners and customers, so some incidents are mitigated externally, including by \cloud{} Partners and customers.
For example, engineers will recommend that customers modify their prompts when their wrong usage causes the model to return unexpected content or switch to a stable model.

\subsection{Self-recover (19.7\%)} 
\label{subsection:self-recover}
These transient incidents are automatically mitigated as the service recovers on its own due to its resilience mechanisms, for example, back-off retry, or when the monitoring system no longer detects abnormal indicators, e.g., heartbeat detection rate returns to normal. 
Note that self-recovered incidents are not false alarms in our dataset.

\section{Discussion}
\subsection{Lessons Learned}
\label{subsec:lessons}

\begin{figure}[t]
  \centering
   \includegraphics[width=\linewidth]{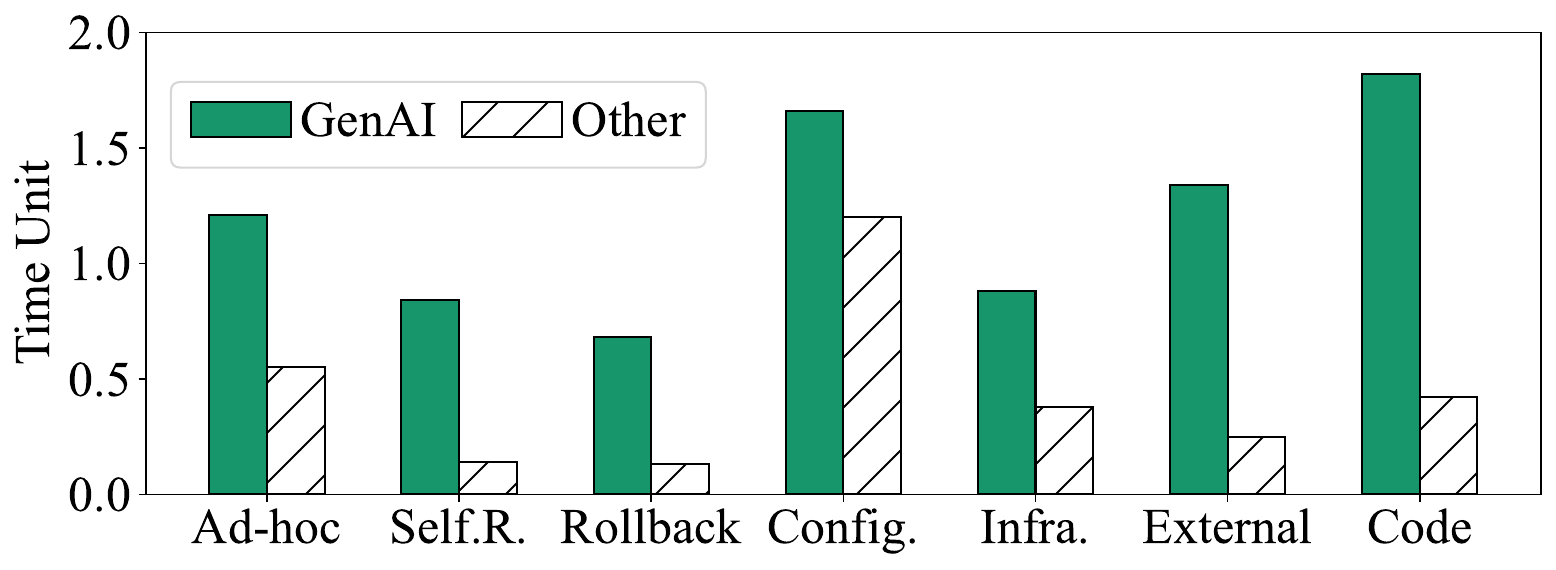}
  \caption{Average TTM for different mitigation approaches. 
  }
   \label{fig:ttm_avg}
·\end{figure}

Since the mitigation strategy categories for both \gai{} and non-\gai{} share high similarities, we further perform a comparative study to identify their distinctions. We find that \gai{} incidents generally require more time to mitigate compared to other types.
Specifically, on average, \gai{} incidents take 1.12 time units to resolve, compared to 0.65 time units for non-\gai{} incidents.

To reveal the underlying reason:
(1) We calculate the TTM for each type of mitigation category, and find that the longer TTM for \gai{} incident holds across all mitigation categories, as shown in \autoref{fig:ttm_avg}, reflecting the complexity of solving various \gai{} incidents.
Additionally, across all factors we consider (severity levels, detection types, troubleshooting guides) in the general analysis in \autoref{subsec:LLM_mitigate}, the Time to Mitigation (TTM) for LLM incidents is consistently longer than for incidents in other services.
(2) We compare the distribution of mitigation approaches, as depicted in
\autoref{fig:mitigation}. The ad-hoc fix (54.7\%) is the majority of the mitigation for other cloud services, which have shorter TTM compared to any \gai{} incident mitigation in \autoref{fig:ttm_avg}. The mitigation distribution of \gai{} incidents is more balanced, with ad-hoc fixes comprising only 22 4\%.
This indicates that, for \gai{} cloud services in their early development stage, more diverse,  sophisticated, and time-consuming methods are required  
as opposed to applying the ad-hoc fixes.
(3) The current monitoring tools for \gai{}  cloud services are being continuously improved to better align with their unique requirements. Enhancements in accuracy and adequacy are expected to help reduce TTM and improve overall efficiency.
Unlike conventional cloud services monitored by automated watchdogs, a high percentage of \gai{} incidents are detected by humans. According to \autoref{table:false_alarm_and_detection} in \autoref{subsec:llm_detection}, only 13.7\% of the incidents were detected by humans for non-\gai{} cloud services in our dataset, compared to 38.3\% for \gai{} incidents. 
Furthermore, monitor-detected \gai{} incidents have an 11.0\% false positive alarm rate, significantly higher than the 3.8\% observed in other services. 
This suggests that the current monitor is not mature compared to conventional incidents, and requires additional effort to improve. 

Longer TTM is also attributed to the difficulty in performing root cause analysis for \gai{} incidents. As discussed in \autoref{sec:mitigation}, a single symptom can stem from multiple root causes, thus complicating the debugging of \gai{} services. For example, diagnosing unexpected model outputs can be complex; potential causes include faulty hardware, misconfigurations, code defects, or misuse.

\subsection{Implications}
Our findings offer actionable insights for a wide range of stakeholders, including researchers, model providers, service maintainers, developers, and etc.

\point{Researchers} Our study highlights several avenues for future research, particularly in automated methods to detect invalid inference results. 
Currently, invalid outputs (14.5\%), such as hallucinations or irrelevant responses, are challenging to detect. The current state-of-the-art detection methods generally include 1) self-judgment by the LLM, 2) fine-tuning another model with human-labeled data, or 3) calculating consistency scores after multiple attempts. However, neither of them is cost-efficient nor fully effective. More robust research is needed to address these limitations and develop scalable validation algorithms that can operate across various \gai{} applications. 

\point{Model Providers} Besides the high ratio of invalid inference results (14.5\%) and challenges in detecting hallucinations or invalid content, another notable finding is that 38\% of \gai{} incidents are reported by humans, reflecting that monitoring tools are underdeveloped.
Moreover, many \gai{} cloud services (45.9\%) are still under development or in the preview stage, coupled with the scarcity of incident monitor types. 
Providers should enhance service observability to detect and diagnose issues more effectively, and provide better support and documentation to help users navigate the complexities of \gai{} service integration and management.

\point{Service Maintainers} Our study reveals that the Time-to-Mitigate (TTM) for \gai{} incidents is 1.83 times longer than for non-\gai{} incidents, highlighting the need for automation in incident mitigation. The complexity of \gai{} systems, which involve vast and interconnected layers of infrastructure, dependencies, and configurations, is a significant factor. For example, \gai{} cloud systems require 2.5x more infrastructure fixes, 3.0x more code changes, and 3.0x as many configuration updates compared to non-\gai{} services. Despite these, more straightforward ad-hoc fixes are applied in only 22.4\% of \gai{} incidents, compared to 54.7\% in non-\gai{} services, indicating a reliance on more complex, time-consuming fixes for \gai{} systems. 
Furthermore, diagnosing root causes of \gai{} incidents is often complex. A single symptom, such as poor performance (49.8\%) or deployment failure (35.7\%), can have multiple root causes, including infrastructure problems (27.2\%), configuration problems (24.5\%), or code bugs (22.5\%). Services should provide observability from different dimensions to obtain granular insight into these symptoms and their underlying causes.
Maintainers should consider 1) implementing more automation tools or agents for distinct mitigation approaches, 2) adopting more infrastructure-as-code practices to manage complex \gai{} cloud infra more effectively, and 3) integrating more automated rollback mechanisms to address compatibility issues swiftly.

\point{Application Developers and Users} For developers, input validation and dynamic rate limiting are critical areas needing improvement. Incidents reveal that special characters, fragmented prompts, and excessive token usage, even within token limits, can disrupt model processing. Developers should implement input validation processes to prevent these issues and adopt dynamic rate-limiting strategies.

\section{Related Work}
\point{Empirical studies on cloud incidents} 
A significant amount of prior work has been devoted to studying the characteristics of incidents occurring in production systems. Ganatra et al. \cite{detection_better_cur} examined incident detection at Microsoft to identify monitoring gaps in cloud platforms. Chen et al. \cite{incident_triage_oss} studied incident triage in Microsoft's online service systems to understand industry practices. Zhao et al. \cite{changed_incidents} explored change-induced incident lifecycles in large-scale online services, offering management insights. Wang et al. \cite{how_long_to_mitigation} analyzed the time-to-mitigation (TTM) of incidents across 20 Microsoft online services. Building on this, our study delves into incident characteristics, comparing incidents related to \gai{} with those of other services. In related work, Liu et al. \cite{bugs_make_production} investigated software bugs causing cloud incidents in Microsoft Azure and their resolutions. Ghosh et al. \cite{how_to_fight} analyzed incidents in Microsoft Teams, classifying root causes and mitigation steps. Martino et al. \cite{sass_failures} characterized failures in a business data processing platform using event log data.
Our work builds upon these studies by applying established approaches from traditional cloud incident analysis to the \gai{} cloud services context. While following similar research methodologies, we highlight characteristics unique to \gai{} incidents, such as symptoms that involve invalid inference, which are not commonly seen in conventional cloud environments.

\point{LLMs empirical study} In recent years, with the rise of large language models (LLMs), numerous related studies have emerged. Cui et al.~\cite{cui2024risk} organize existing studies related to LLMs and propose a comprehensive taxonomy, which systematically analyzes potential risks in LLM systems and discusses corresponding mitigation strategies. Liu et al.~\cite{liu2305jailbreaking} investigate the use of jailbreak prompts to bypass restrictions imposed on ChatGPT. They conduct an empirical study to evaluate the effectiveness and robustness of prompts collected from the real world. Zhuo et al.~\cite{zhuo2023robustness} present an empirical study on the adversarial robustness of a prompt-based semantic parser based on Codex. Yang et al.~\cite{yang2022empirical} conduct a study on GPT-3 in knowledge-based visual question answering (VQA), treating GPT-3 as a knowledge base (KB) and adapting GPT-3 to solve the VQA task in a few-shot manner.
In contrast to these studies, which primarily focus on model behavior and robustness, our work centers on the reliability of LLM-related cloud services. Specifically, we present a novel empirical study of incidents in such services, offering insights into their design, operation, and maintenance.

\section{Threats to Validity}

\point{Internal threat} 
Subjectivity may occur during manual labeling as an internal threat. To mitigate this threat, our study go through multiple rounds involving independent labeling, meetings to discuss categorization, and the calculation of Cohen’s kappa~\cite{kappa}. We ultimately select the round of labeling that is near-perfect as our final result, which demonstrates the highest consistency.

\point{External threat} All incidents we collect come from \cloud's cloud systems. Given that \cloud{} employs various effective tools and techniques to eliminate bugs and deploys multiple automated tools to mitigate some incidents before they impact customers, the incidents we collect may not fully represent the behavior of other \gai{} cloud services. 
We plan to perform a larger scale evaluation of \gai{} cloud services from different companies in the future.

\section{Conclusion}
In this paper, we present a comprehensive study of incidents from \gai{} cloud services within \cloud. We explore the symptoms, root causes, and mitigation strategies of \gai{} incidents.
Our findings reveal unique characteristics in \gai{} cloud services.
For example, we identify notable differences between incidents from LLM cloud services and other cloud services, such as significant disparities in the time to mitigation of incidents. 
Additionally, we find that the primary cause of incidents in LLM cloud services is related to infrastructure. 
These findings provide guidance for future academic and industrial research in the field of LLM cloud services.
We hope to inspire the development of advanced, specialized tooling and raise discussions on \gai{} incidents, so that our community can monitor the \gai{} cloud system with early warnings, triage incidents to the correct teams with fewer hops, pinpoint root causes accurately, and mitigate the incidents with optimal plans.

\section*{Acknowledgement}
We sincerely thank all anonymous reviewers for their valuable feedback and guidance in improving this paper. This work was sponsored by National Natural Science Foundation of China (No.62372193 and No.U2436207).

\bibliographystyle{ieeetr}
\bibliography{ref}

\end{document}